 \documentclass{pasj00}

 \SetRunningHead{H.~Imai \etal}
{CO $J=3-2$ emission from the water fountain sources}
 \Received{2009/05/??}
 \Accepted{2009/??/??}
 \def \vlsr{$V_{\mbox{\scriptsize LSR}}$}
 \def \kms{~km~s$^{-1}$}
 \def \etal{~et~al.}
 \def \h2o{H$_{2}$O}
 \def \i1634{IRAS~16342$-$3814}
 \def \j1828{IRAS~18286$-$0959}
 \def \vjet{$V_{\mbox{\scriptsize jet}}$}

 \title
{CO $J=3-2$ emission from the ``water fountain" sources  \i1634 \ and  \j1828}

 \author{Hiroshi  \textsc{Imai} \altaffilmark{1}, 
Jin-Hua  \textsc{He} \altaffilmark{2}, 
Jun-ichi  \textsc{Nakashima} \altaffilmark{3}, 
Nobuharu  \textsc{Ukita} \altaffilmark{4}, 
Shuji  \textsc{Deguchi} \altaffilmark{5}, \\
and 
Nico  \textsc{Koning} \altaffilmark{6}
}

 \altaffiltext{1}{Graduate School of Science and Engineering, 
Kagoshima University,  \\
1-21-35 Korimoto, Kagoshima 890-0065}
 \email{hiroimai@sci.kagoshima-u.ac.jp}

 \altaffiltext{2}{National Astronomical Observatories/Yunnan Observatory, 
Chinese Academy of Sciences,  \\
PO Box 110, Kunming, Yunnan Province 650011, China}

 \altaffiltext{3}{Department of Physics, University of Hong Kong, Pokfulam Road, 
Hong Kong, China}

 \altaffiltext{4}{ALMA Project Office, National Astronomical Observatory,  
2-21-1 Osawa, Mitaka, Tokyo 181-8588} 

 \altaffiltext{5}{Nobeyama Radio Observatory, National Astronomical Observatory, 
Minamimaki, Minamisaku, Nagano 384-1305}

 \altaffiltext{6}{Department of Physics and Astronomy, University of Calgary, Calgary, AB T2N 1N4, Canada}


\begin{document}

 \maketitle


 \begin{abstract}

We observed CO $J=3-2$ emission from the ``water fountain" sources, which exhibit high-velocity collimated stellar jets traced by  \h2o maser emission, with the Atacama Submillimeter Telescope Experiment (ASTE) 10~m telescope. We detected the CO emission from two sources,  \i1634 and  \j1828. The  \i1634 CO emission exhibits a spectrum that is well fit to a Gaussian profile, rather than to a parabolic profile, with a velocity width (FWHM) of $158 \pm 6$ \kms \ and an intensity peak at  \vlsr$=50 \pm2$ \kms. The mass loss rate of the star is estimated to be $  \sim 2.9\times 10^{-5}M_{ \odot}$~yr$^{-1}$. Our morpho-kinematic models suggest that the CO emission is optically thin and associated with a bipolar outflow rather than with a (cold and relatively small) torus. The  \j1828 CO emission has a velocity width (FWHM) of $3.0 \pm 0.2$ \kms, smaller than typically seen in AGB envelopes. The narrow velocity width of the CO emission suggests that it originates from either an interstellar molecular cloud or a slowly-rotating circumstellar envelope that harbors the \h2o maser source. 
 \end{abstract}


 \section{Introduction}

The process in which mass outflow shapes planetary nebulae (PNe) is still unresolved and needs to be elucidated. One of the striking features of some PNe is the bipolarity in their morphological and kinematical structures, some of which clearly exhibit high velocity bipolar jets ($V_{\mbox jet}>$100 \kms). Such bipolar jets have been found at the beginning of the post asymptotic giant branch (AGB) phase and even at the end of the AGB phase (e.g. \cite{sah98,ima02,ima07b}). In particular, a small fraction of stellar jets have velocities higher than a typical expansion velocity of a circumstellar envelope (CSE) found in 1612~MHz OH maser emission (\vjet $ \gtrsim$30\kms), and are traced by  \h2o \ maser emission. Such high velocity jets are called ``water fountains" \citep{ima07b}. 
As of 2009, 13 such objects have been identified \citep{wal09,sua09}. Some spatio-kinematical structures of the jets have been elucidated by very long baseline interferometry (VLBI) observations of the  \h2o masers  \citep{ima02,ima05,ima07,ima07b, bob07,cla09}.  Interestingly, all of the estimated apparent dynamical ages of the jets are about 100~yr or shorter, which is consistent with the rare detection of water fountain sources. This implies that these jets should have just been launched at the final evolutionary stage of the dying stars. Note that the  \h2o maser emission is excited when the bipolar tips of the stellar jet strike into the ambient CSE that was produced from the previous spherically symmetric stellar mass loss. Therefore, in order to more reliably estimate the stellar mass loss rate, the dynamical age, and the whole morphological and kinematical structure of the jet, a mapping observation of thermal emission such as CO emission is   crucial. 

The first detection of CO emission from a water fountain source was reported with the Arizona Radio Observatory 10~m telescope towards  \i1634  \citep{he08}. Usually it is difficult to detect such CO emission towards the water fountain sources because they are located close to the Galactic plane with heavy contamination from the interstellar CO emission or they are too distant ($D \gtrsim$2~kpc). Here we present results of the first systematic CO $J=3-2$ emission observations towards the water fountain sources with the Atacama Submillimeter Telescope Experiment (ASTE) 10~m telescope. In addition, we also show results of CO $J=1-0$ emission observations towards W~43A and  \j1828 with the Nobeyama 45~m telescope. In section 2 and 3, we describe in detail the observations and results, respectively. In section 4, we discuss the mass loss rate and the morphology and kinematics of  \i1634, whose CO $J=3-2$ emission is detected.


 \section{Observations and data reduction}

\subsection{ASTE observations}
The ASTE observations of $^{12}$CO $J=3-2$ emission at 345.79599~GHz were made on June 20 and 21, 2008 during LST 13:00--22:00. The FWHM beam size of the ASTE telescope is 22\arcsec\ at the observed frequency. For antenna temperature calibration, the CO emission from IRC$+$10216 was observed in a 10~min scan on both days. Comparing the antenna temperature of IRC$+$10216 with that obtained with the Caltech Submillimeter Observatory (CSO) 10~m telescope, we obtained an antenna aperture efficiency $  \eta_{ \mbox{ \scriptsize{MB}}}=$0.59 with reasonable agreement with those found in previous ASTE observations. An antenna pointing check was made every $  \sim$2~hr. The antenna pointing was confirmed to be stable within $  \sim$5\arcsec. The system temperature was between 200 and 400~K (single side band). The received signals were down-converted in frequency and transferred into three base band channels (BBCs), each of which has a band width of 512~MHz, corresponding to a velocity width of 445 \kms \ at 345~GHz. The center frequencies of the BBCs were split by 100~MHz  to check the intrinsic CO emission, which should be detected in different spectral channels but at the same velocity ranges in the all BBCs. We used the MAC spectrometer to obtain a spectrum with 1024 spectral channels, corresponding to a frequency and velocity spacings of 500~kHz and 0.45 \kms, respectively. Reduction of the spectral data was made using the NewStar package developed at the Nobeyama Radio Observatory. 

Table  \ref{tab:water-fountains} gives parameters of the water fountain sources and the observations. The observations were made in the antenna-position switching mode; the number of points observed on and around the target source ($>$1 for the cross-scan mode) is given in Column 10. Because  \j1828, W~43A, and IRAS~18460$-$0151 are located very close to the Galactic plane with intense background molecular emission, five-points or nine-points cross scans were adopted, in which the observed points were separated by 10\arcsec --20\arcsec (see sect.  \ref{sec:results}). Column 11 in table  \ref{tab:water-fountains} gives root-mean-square noise levels of the spectra. Note that the raw spectrum after integration had an emission-free baseline with a linear gradient. This situation was nearly ideal for detecting any velocity component of CO emission. The linear gradient baseline was subtracted to obtain the final spectrum. 

\subsection{Nobeyama 45~m observations}
The Nobeyama 45~m observations of $^{12}$CO $J=1-0$ emission at 115.271204~GHz were made towards W~43A on December 1--2, 2008 during LST 16:00--21:00 and towards \j1828 on April 29, 2008 with a beam size of about 15\arcsec. A single beam SIS receiver and the $5\times 5$ beam focal plane array receiver system (BEARS, \cite{sun00}) were used to make five point and 25 point scans around the target sources, respectively. The grid spacing of BEARS was 41\arcsec, while that of the five point scans was set to 30\arcsec. The system temperatures were between 400 and 600~K (single side bands). To observe W~43A and \j1828, both of the two receiving systems and BEARS were used, respectively. For signals received with the SIS receiver and BEARS, the acousto-optic spectrometers (AOSs) and the digital spectrometers \citep{sor00} were used, respectively, to obtain spectra that have 2048 and 1024 spectral channels, yielding velocity spacings of 0.33\kms\ and 0.16\kms(in a two-channel binding mode) each at 115~GHz, respectively. In order to find intrinsic emission from the target sources, the spectra obtained from four points around each target were averaged and the average spectrum was subtracted from that obtained at the target position. 


 \section{Results}
 \label{sec:results}

We found CO $J=3-2$ emission with ASTE towards  IRAS~15445$-$5449, IRAS~15544$-$5332, \i1634,  \j1828, W~43A, IRAS~18460$-$0151, 
and IRAS~18596$+$0315. From  IRAS~15445$-$5449, IRAS~15544$-$5332, W~43A, IRAS~18460$-$0151, and IRAS~18596$+$0315, however, the emission must originate from background sources because no intrinsic CO emission is detected. If it indeed exists, it should be detected only at the exact position and systemic velocity of the source. It has been observationally found that the center velocities of \h2o\ and 1612~MHz OH 
maser spectra in the water fountain sources are coincident with each other to within 1--2\kms \citep{ima02,ima07b}. This implies that the systemic velocities of the driving sources of these masers as well as the CO envelope/flow should coincide within the above range.  The systemic velocities are cited from previous \h2o \ and SiO maser spectra (see table \ref{tab:water-fountains}). At positions around W~43A, the same velocity components with equal antenna temperature in each component are found. Separately, we have observed $^{12}$CO, $^{13}$CO $J=1-0$ and HCN emission towards W~43A with the Nobeyama 45~m telescope, but obtained negative detection (with upper limits of 0.2~K, 0.1~K in antenna temperatures for $^{12}$CO and $^{13}$CO $J=1-0$ lines, respectively). The $J=3-2$ CO emission towards IRAS18286-0959 and IRAS16342-3814 is detected at the precise position and velocity of each source. Towards other water fountain sources, no CO emission was detected in the position-switching observations. 

\begin{figure}
     \FigureFile(85mm,85mm){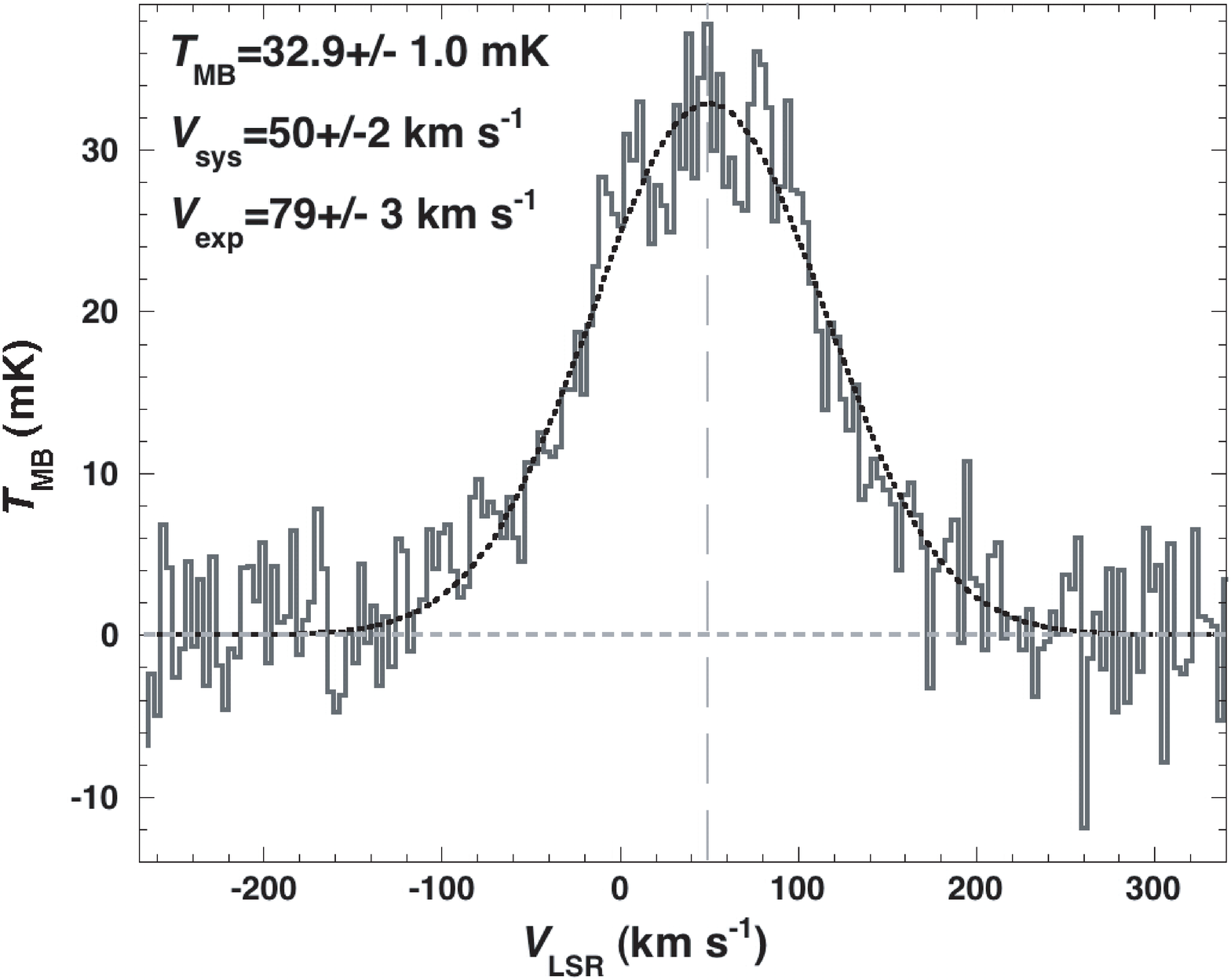}
    \caption{CO $J=3-2$ emission spectrum of  \i1634. The horizontal grey dashed line shows the zero-temperature baseline. The dotted line shows the Gaussian fit  with a peak of $T_{ \mbox{\scriptsize MB}}=32.9 \pm1.0$~mK at LSR velocity of 50$  \pm$2 \kms (a vertical dashed line) and a HWHM width of 79$  \pm$3 \kms.} \label{fig:I16342}
 \end{figure}

Fig.  \ref{fig:I16342} shows the CO $J=3-2$ spectrum towards \i1634 (hereafter abbreviated as I16342). To improve the sensitivity, three MAC spectra covering different velocity ranges are synthesized although the intrinsic CO emission is already identified in each of the spectra. The synthesized spectrum is well fit to a Gaussian profile with a velocity width (FWHM) of $158 \pm 6$ \kms \ and a peak intensity of $I_{\mbox{\scriptsize peak}}=32.9\pm1.0$~mK at  \vlsr$=50 \pm2$ \kms\ with respect to the local standard of rest (LSR). Remarkably, the velocity width is larger than that of 1612~MHz OH maser emission ($\sim$120\kms, \cite{sah99,cla09}) and comparable to the full width of the  \h2o maser spectrum (e.g., \cite{lik92,cla09}). The origin of the wide wing of the spectrum is discussed in sect. \ref{sec:morphology}. 

\begin{figure*}
     \FigureFile(170mm,170mm){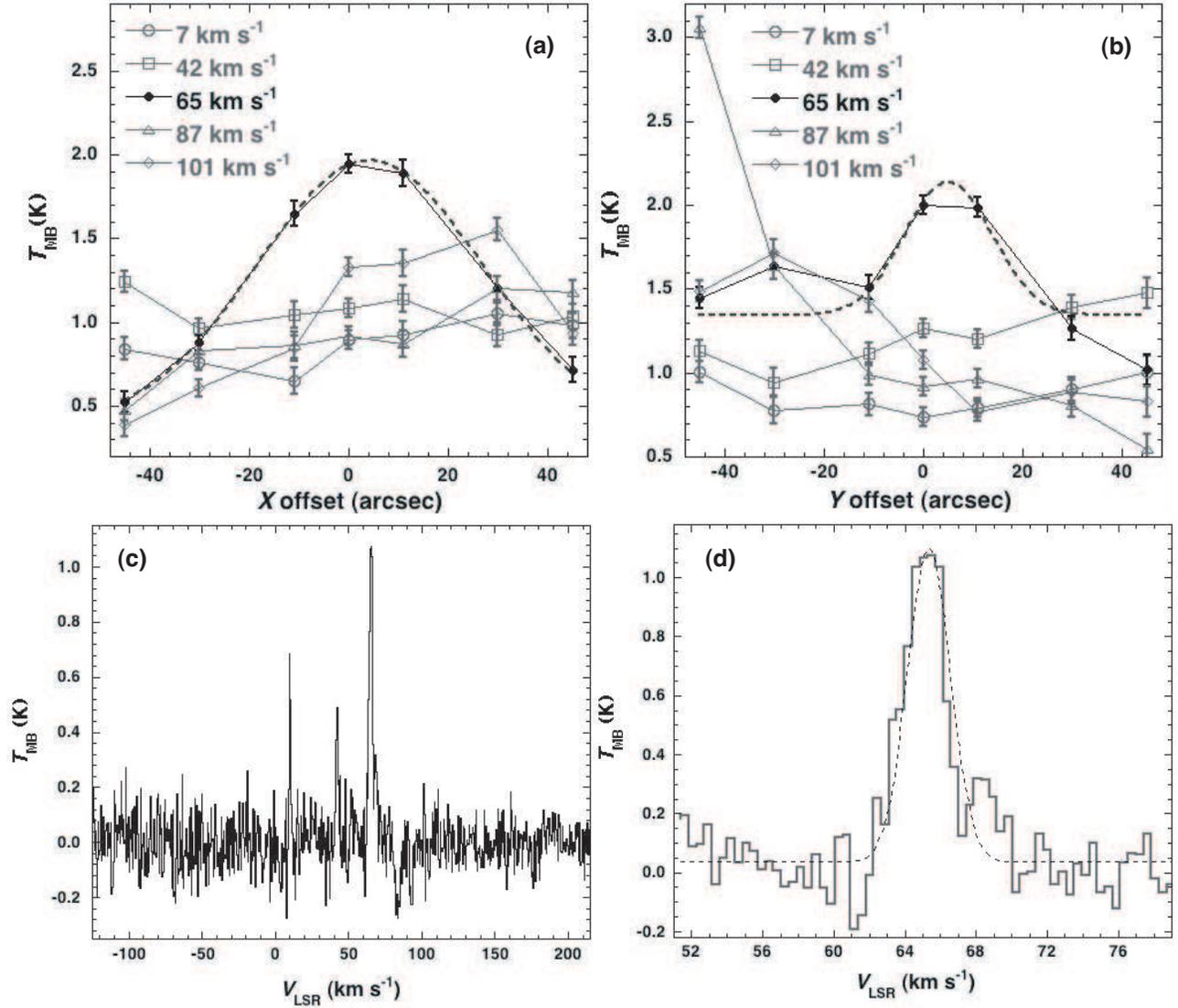}
 \caption{CO $J=3-2$ emission in  \j1828.  (a) Temperature variations of CO emission peaks along the position offset from  \j1828 in the east--west ($X$) direction. The dashed line shows the Gaussian fit for the 65 \kms\ peak with a peak of $T_{ \mbox{\scriptsize MB}}=1.58 \pm0.03$~K at a peak offset $X_{0}=$4 \farcs3$  \pm$0 \farcs2 and a HWHM width of 27 \farcs3$  \pm$0 \farcs6. (b) Same as (a) but in the north--south ($Y$) direction. The dashed line shows the Gaussian fit  for the 55 \kms\ peak with a peak of $T_{ \mbox{MB}}=0.79.8 \pm0.33$~K at a peak offset of $Y_{0}=$5 \farcs1$  \pm$3 \farcs8 and a HWHM width of 10 \farcs6$  \pm$7 \farcs3. (c) Spectrum at the position of  \j1828, which is subtracted by the mean spectrum synthesized from  spectra at four positions around  \j1828. (d) Same as (c) but enlarged in velocity around the emission peak. The dashed line shows the Gaussian fitting slope with a baseline offset of 0.04$ \pm$0.01~K, a peak of $T_{ \mbox{\scriptsize MB}}=1.07 \pm0.06$~K at LSR velocity of 65.3$  \pm$0.1 \kms, and a FWHM width of 3.0$  \pm$0.2 \kms.}
 \label{fig:I18286}
 \end{figure*}

\begin{figure}
    \FigureFile(85mm,85mm){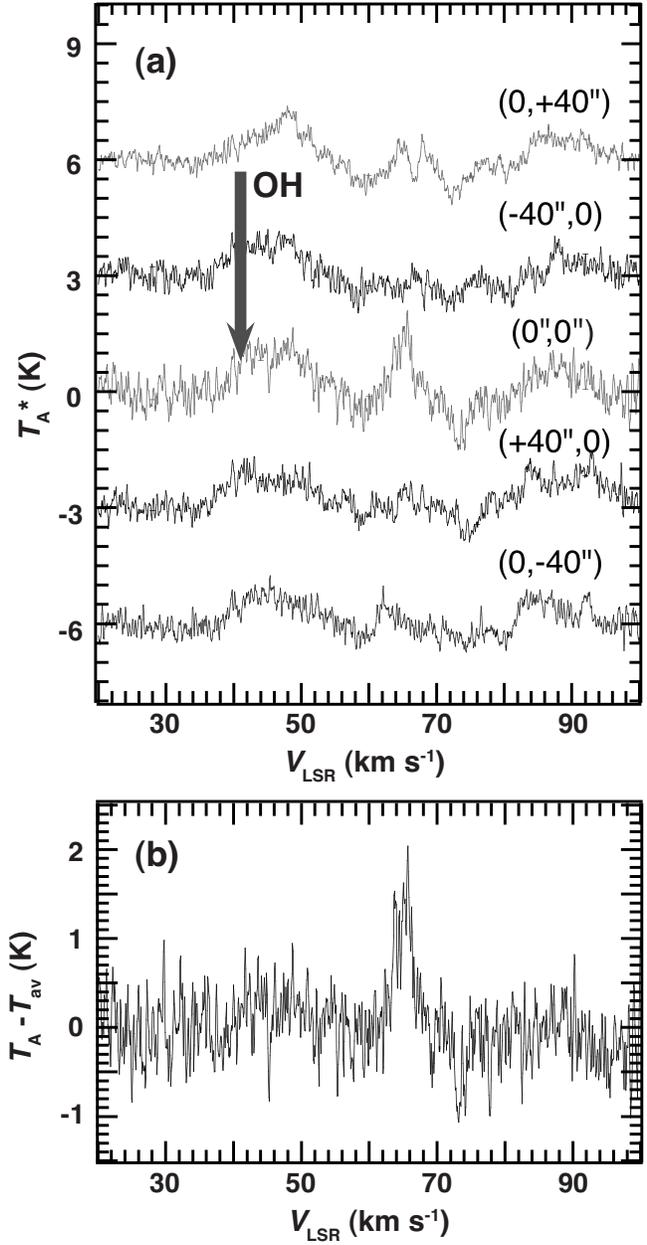}
 \caption{Spectra of CO $J=1-0$ emission towards  \j1828 and the adjacent points observed with the Nobeyama 45~m telescope. (a) Raw spectra at \j1828 and four adjacent points, whose relative positions are described at the right sides of the spectra. The antenna temperature values of the spectra are shifted in a step of 3 K for clarity. There exists a narrow peak at  \vlsr $  \simeq$65 \kms, which is brightest at  (0\arcsec, 0\arcsec) and expected to be the intrinsic emission from  \j1828.  Two wide bumps at \vlsr $  \simeq$50 \kms\ and \vlsr $  \simeq$90 \kms\ may be the Galactic background CO emission. (b) Spectrum towards \j1828 obtained after an average spectrum synthesized from the adjacent points is subtracted.} 
 \label{fig:NRO45m}
 \end{figure}

\h2o masers in \j1828 were discovered by \citet{deg07}. Negative detections of HCO$^{+}$, H$\alpha$53, and NH$_{3}$ lines toward this object suggest that they are not associated with any dense molecular cloud or compact H{\rm II} region. Although a plenty of velocity components in the \h2o masers may imply their association of a massive young stellar object, their spatio-kinematical structure clearly exhibits collimated jets as seen in water fountain sources \citep{ima07b}. Fig.  \ref{fig:I18286} shows the result of the CO $J=3-2$ observation towards \j1828. Towards \j1828 and nearby positions, five velocity components are found in the CO $J=3-2$ spectra. Only the velocity component at  \vlsr=65\kms\ has its antenna temperature maximum at the position of \j1828 (fig.  \ref{fig:I18286}a and b).  An off-source spectrum was synthesized from the spectra at four positions around this source, and was subtracted from the on-source spectrum. In the five velocity components, only one velocity component at \vlsr$\simeq$65\kms\ was identified as possible intrinsic emission from \j1828 (Fig.  \ref{fig:I18286}c). As shown in figure \ref{fig:NRO45m}, the same velocity component is also found in CO $J=1-0$ emission. However, this component has a velocity width (FWHM) of only 3\kms (fig.  \ref{fig:I18286}d), which is smaller than those typically found in AGB envelopes  ($\sim$10\kms, e.g. \cite{kem03}) and much smaller than the velocity range of the  \h2o\ maser emission in the same source  \citep{deg07,ima07b}. Our unpublished European VLBI Network (EVN) data shows that 1612~MHz OH maser emission is detected at \vlsr=39.5\kms\ towards \j1828. We believe that the OH maser component is one of double-peaked components as typically seen in stellar 1612~MHz OH maser sources. If the CO and OH lines are generated from the common circumstellar envelope, a terminal expansion velocity is derived to be $\sim$25\kms, which is inconsistent with the narrow line width of the CO emission. Using Equation \ref{eq:mass-loss} in sect.\ \ref{sec:mass-loss}, we derive a mass loss rate to be $\sim 3.5\times 10^{-7}\:M_{\odot}$~yr$^{-1}$, inconsistent with the water fountain sources being at the final AGB or early post-AGB phase. We think that the CO emission may be associated with a background young stellar object 
seen in the same direction. Alternatively, it may be associated with a rotating circumstellar disk surrounding the central object that collimates the \h2o maser flow. In fact, the existence of such circumstellar disks has already been proposed (e.g., \cite{jur99,nak06}). In order to unambiguously elucidate the origin of the CO $J=3-2$ and $J=1-0$ lines, mapping observations with radio interferometers are indispensable.


 \section{Discussion}

 \subsection{Stellar mass loss rate of  \i1634}

 \label{sec:mass-loss}

In the present paper, we attempt to estimate a mass loss rate of the I16342 CO $J=3-2$ outflow. In spite of the limited quality of the $^{13}$CO $J=2-1$ spectrum,  \citet{he08} suggests that the CO $J=2-1$ line is optically thin on the basis of a low 
$^{12}$CO antenna temperature. In addition, the CO $J=3-2$ line is axisymmetric with respect to the systemic velocity. Thus 
it is expected that the CO $J=3-2$ line is optically thin. In this case, using Eq. (6) and (8) of  \citet{kna85} and a correction for the beam size and the different CO transition (cf. \cite{olo93,gro99}), a mass loss rate is estimated in units of solar masses per year as follows, 

 \begin{equation}
 \dot{M}_{\mbox{\scriptsize{gas}}}=4.55\times 10^{-19}
 \left[\frac{T_{\mbox{\scriptsize MB}}}{\mbox{Log}(W/0.04)s(J)}\right]^{5/6}f^{-1}_{\mbox{\scriptsize CO}}
 V^{11/6}_{\mbox{\scriptsize exp}}(DB)^{5/3}. 
\label{eq:mass-loss}
 \end{equation}

 \noindent
Here $T_{\mbox{\scriptsize{MB}}}$ is the antenna temperature of the CO emission in Kelvin, $V_{\mbox{\scriptsize exp}}$ the expansion velocity of the CO emission in \kms, $D$ the source distance in parsec, $B$ the beam size of the telescope, $f_{\mbox{\scriptsize{CO}}}$ the abundance of CO molecules relative to H$_{2}$, $s(J)$ a correction factor for $J\rightarrow J-1$ transition, and $W$ the ratio of the 4.6~$\mu$m flux to that emitted by a blackbody of temperature of 2000~K and radius of 5$\times 10^{13}$~cm 
(i.e., $W$=1). For  I16342, $T_{\mbox{\scriptsize{MB}}}=$33~mK, $D=$2~kpc, and $B=$22\arcsec\ are adopted. We also adopt a terminal velocity projected on the line of sight, $V_{\mbox{\scriptsize exp}}=$120\kms (e.g, \cite{lik92,sah99,cla09}). For an O-rich circumstellar envelope harboring  \h2o \ and OH maser emission, $f_{\mbox{\scriptsize{CO}}}=5 \times 10^{-4}$ is adopted. We assume $s(3)=$0.43 on the basis of previous observations with the CSO 10~m telescope  \citep{kna98,gro99}. The unknown factor $W$ (0.1$\lesssim W\lesssim$5) introduces an uncertainty of the mass loss rate within a factor of 0.4--2.1. 

Thus we derived the mass loss rate of  I16342, $  \dot{M}_{\mbox{\scriptsize{gas}}} \approx 2.9\times10^{-5} \:M_{\odot}$~yr$^{-1}$. If we adopt a HWHM of the CO spectrum to be the terminal expansion velocity, $V_{\mbox{\scriptsize exp}}=$79\kms, then we obtain a value $  \dot{M}_{\mbox{\scriptsize{gas}}} \approx 1.3\times10^{-5} \:M_{\odot}$~yr$^{-1}$, comparable to that previously estimated with CO $J=$2--1 emission ($  \dot{M}_{\mbox{\scriptsize{gas}}} \approx 1.7\times10^{-5} \:M_{\odot}$~yr$^{-1}$ obtained after correction\footnote
{The previous calculation adopted a beam size of the IRAM 30~m telescope to be 21\arcsec, but which should be corrected to be 12\arcsec. When the derived mass loss rate was scaled to that of the Arizona Radio Observatory 10~m telescope (32\arcsec), we obtained the new value for the CO $J=2-1$ emission.}, \cite{he08}). Note that these calculations assume spherically symmetric mass loss. As mentioned later, because the CO emission may be associated with a collimated, fast outflow, the mass loss rate is expected to be overestimated. 
Note also that the derived mass loss rate is much lower than 
roughly estimated with mid-IR emission ($ \dot{M}_{ \mbox{\scriptsize{gas}}}  \approx  10^{-3} \:M_{\odot}$~yr$^{-1}$, \cite{dij03}). Because the latter adopted a much lower expansion velocity (15\kms), leading to a higher mass loss rate with a higher adopted expansion velocity, there exists a large discrepancy between the rate values. \citet{zij01} argues the possibility that the value derived in the latter is attributed to the extreme extinction coming from an edge-on thick disk, which may be invisible in the CO emission as explained later. Using any calculated mass loss rate, it is concluded that  I16342 should be at the phase of stellar evolution, where its mass loss is highest. The success of the CO emission detection towards I16342 may be attributed to its close proximity to the Earth and the large offset from the Galactic plane, which enables us to more easily obtain a long integration time more easily. Even with a similar mass loss rate and a similar source distance, we need a much longer observation time towards W~43A including adjacent off-target positions in the Galactic plane. Alternatively, interferometric observations enable us to obtain higher sensitivity as achieved with the Atacama Large Millimeter and Submillimeter Array (ALMA) and to distinguish intrinsic CO emission from the background emission. 

 \subsection{Copious mass loss from an expanding flow in  \i1634 }

 \label{sec:morphology}

In the case of an optically thick spherically-expanding circumstellar envelope, the observed CO emission shows a parabolic spectral profile, which has been confirmed in many AGB stars (c.f,  \cite{kem03}). When a bipolar outflow is a dominant component of the CO emission, the spectrum exhibits double-sided {\it wings} whose full velocity width indicates twice the velocity of the bipolar flow. Such wing profiles are found in the spectra towards young stellar objects (YSOs) and the spectra themselves resemble Gaussian profiles. More intense CO emission around the systemic velocity suggests that the bipolar outflow is accelerated in the vicinity of a YSO. Even if the YSO is deeply embedded in a dense molecular cloud core, the core itself may not be detected in the higher $J$ transition of CO emission when it is too cold to excite the emission and/or its filling factors are too small in the single-dish beam. The CO spectrum of  I16342 resembles such a Gaussian profile, however it is difficult to expect a physical and dynamical condition similar to those of star forming regions.

In order to better understand the morpho-kinematical structure of the I16342  CO flow, we have constructed models using the {\it Shape} software package (see, e.g., \cite{ste06}). {\it Shape} is a morpho-kinematic modeling tool used to create three-dimensional(3D) models of astronomical nebulae. It has also been used for obtaining models of circumstellar CO envelopes (e.g., \cite{nak09}).  Note that {\it Shape} does not calculate the numerical hydrodynamic evolution, temperature profile, or  radiation transfer equations in detail. Instead, a monte-carlo algorithm is introduced, in which a three-dimensional space is divided into cubes with a specific temperature and contributions to emission, absorption, and scattering from the individual cubes are calculated. Because {\it Shape} has been updated since the calculations in \citet{nak09}, we are now concerned with the morphology including a density profile, kinematics as function of distance from the central star, and opacity, but do not try to reproduce the intensity scale. 

\begin{figure}
     \FigureFile(85mm,85mm){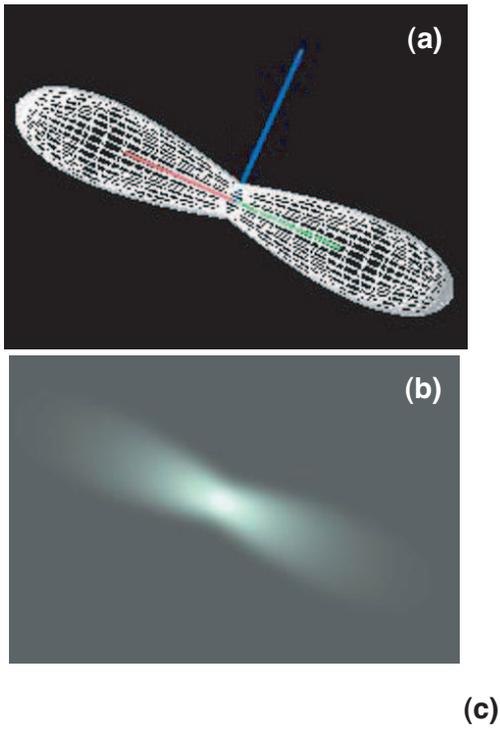}
 \caption{A SHAPE model that reproduces the observed CO spectrum. (a) Mesh image of the modelled morphology. (b) Brightness images of the model. (c) Spectrum obtained from the model (a black line), whose velocity width and peak antenna temperature are scaled to those of the observed spectrum (a grey line).} 
\label{fig:model}
 \end{figure}

Figure  \ref{fig:model} shows one of the best-fit models and its simulated spectrum obtained from an accelerated bipolar flow model. The modelled flow is expanding with a velocity proportional to the distance from the central star and with an exponentially decreasing gas density. The flow is slightly squeezed near the central star in order to reproduce the slightly flat peak of the observed spectrum. The simulated brightness distribution resembles the shape of the Keck optical image \citep{sah05}, except for the central brightest part which is obscured in the latter due to heavy extinction. Note that it is impossible to obtain a unique model to reproduce the observed spectrum without any information on observed spatial brightness distribution. We confirmed that there exists other possible models to reproduce the same Gussian spectral profile, but which are based on different flow morphologies and density profiles. 

Nevertheless, through our  {\it Shape} modelling, we find possible major factors to control the spectral profile and can qualitatively evaluate them. First, as mentioned above, the CO emission should be optically thin. The I16342 CO outflow cannot be approximated to a spherically symmetric flow. Taking into account the flow major axis that may be coincident with the major axis of the \h2o\ maser jet at an inclination of 44 \arcdeg\ with respect to the line of sight and at a position angle of 66 \arcdeg\ east from north \citep{cla09}, an asymmetric profile is easily formed if the CO emission is optically thick. Also taking into account the 3-D speed of 180 \kms \ and the acceleration (or the velocity gradient) of the jet, an optically thin flow is more reasonabl. Secondly, the CO flow should be dominated by a bipoloar flow rather than an expanding torus or an equatorial flow. \citet{nak09} reproduced the spectrum and spatial brightness distribution of the post-AGB star IRAS~07134$+$1005 in term of a torus expanding with a velocity of 8\kms. When adopting a similar morpho-kinematical structure, we find that the simulated spectrum has too high brightness at the high velocity wings. \citet{ver09} proposed the existence of a {\it dusty} equatorial flow that has a biconical cavity along the \h2o\ maser jet axis and a full flow opening angle of 145\arcdeg. Assuming that the CO emission originates from the same equatorial flow, we fail to reproduce the Gaussian wings even by  increasing the flow terminal velocity up to 150\kms. Instead, by assuming the CO emission originating from the cavity of the equatorial flow, we succeed in reproducing the Gaussian wings. This model is also consistent with the velocity distribution found in 1612~MHz OH maser emission, in which the high velocity maser components are significantly spatially deviated from the collimated jet found in  \h2o \ maser emission \citep{sah99,zij01}. Thirdly, a dense torus or an equatorial flow proposed by \citet{dij03} and \citet{ver09}, respectively, may contribute to the emission or self-absorption of the $J=3-2$ CO line from I16342, if it exists. In the equatorial flow model mentioned above, instead of reproducing the Gaussian wings as observed, a bump at the spectrum peak appears. It may be decreased by self-absorption, but this causes an unwanted asymmetry in the profile. A decrease in gas density near the central star also reduces the bump, but this is inconsistent with the existence of a thick torus that is clearly seen in optical and mid-infrared images as a dark lane \citep{sah99,dij03,sah05,ver09}. Here we suppose that, as mentioned above, the CO outflow is slightly squeezed to reduce the contribution of the central part of the modelled CO spectrum (figure \ref{fig:model}a).  The negligible contribution from the dense torus to the observed CO spectrum may be attributed, as mentioned above, to the filling factor of the torus that may be much smaller than that of the CO outflow and/or to the too cold ($\lesssim$20~K). The large difference in the filling factors is expected if the total length of the CO flow is larger than that of the \h2o\ maser jet and the optical lobes ($\sim$3\arcsec, \cite{sah05,cla09}) and/or if the radius of the equatorial flow is limited within 1\arcsec\  \citep{ver09}. For the temperature structure, \citet{he08} demonstrated the existence of warmed walls of the bipolar lobes for generating a wing component of CO. \citet{ver09} also confirmed the extreme extinction should be attributed to cooler dust around the equatorial flow. {\bf Note that, the mass loss rate calucation in equation (1) assumes a constant expansion velocity of the flow while the model adopts a linear acceleration of the flow. 
Even for such a linearly accelerating flow, a mass loss rate can be defined if the density profile has a form of $\rho \propto r^{-3}$. In the model, the outflow density decreases exponentially with distance, leading to a good approximation to the $r^{-3}$ profile.} 
   
Thus, we have obtained several insights into the high velocity CO outflow in I16342 through the {\it SHAPE} modelling. When we obtain radio interferometric images as obtained with the SMA and ALMA, we will be able to obtain a unique morpho-kinematical model with more precise information on the gas density structure to completely explain the observed outflow. This should also be the first step to understanding the process of the (almost) simultaneous developments of spherically-expanding, slow tori and highly-collimated, fast jets as well as their decays in the later evolutionary phase. It has been demonstrated that a developed torus as seen in the I16342 dark lane should be created with a highly collimated jet either almost simultaneously or before the jet launch with a very short time lag  ($\lesssim$300~yr, \cite{hug07}). In fact, for I16342, dynamical time scales of the torus and the jet are estimated to be 320~yr (1000~AU/15\kms) and 110~yr (3000~AU/120\kms), respectively. The evolution of the expanding torus should be explored on the basis of a larger number of sampled sources in different evolutionary phases. I16342 should be an important sample for statistical analyses as demonstrated by \citet{hug07}. Some discrete mass ejection events such as companion interactions are also expected in order to explain the simultaneous evolution of jet and torus. It is also worthwile to note that the CO emission in I16342 originates from a different component than that of IRAS~07134$+$1005, which may be in a later evolutionary status than I16342 in the post-AGB phase \citep{nak09}. Not only water fountain sources but also \h2o\ maser sources associated with planetary nebulae should be the targets for the radio interferometric observations in order to discriminate possible scenarios of the evolution and de-evolution of the stellar jets and the circumstellar tori.  

 \section{Conclusions}

Using ASTE, we have surveyed CO $J=3-2$ emission from nine water fountain sources and detected the CO emission towards  I16342 and  \j1828. I16342 may have a mass loss rate of $ \dot{M}_{ \mbox{ \scriptsize{gas}}}  \approx  5.6\times10^{-5} \:M_{ \odot}$~yr$^{-1}$, with most emission attributed to a bipolar outflow rather than a high velocity expanding torus or a biconical flow with a wide opening angle. On the other hand, to further clarify the intrinsic CO detection from \j1828 with a very narrow velocity width ($\sim 3$\kms), we need interferometric observations of this object. Because the water fountain sources are located at large distances, higher sensitivity is essential in future observations. 

 \bigskip

We deeply appreciate members of the ASTE team for their careful observations, preparation and kind operation support. The ASTE project is driven by Nobeyama Radio Observatory (NRO), a branch of National Astronomical Observatory of Japan (NAOJ), in collaboration with University of Chile, and Japanese institutes including University of Tokyo, Nagoya University, Osaka Prefecture University, Ibaraki University, and Hokkaido University. Observations with ASTE were in part carried out remotely from Japan by using NTT's GEMnet2 and its partner R\&E (Research and Education) networks, which are based on AccessNova collaboration of University of Chile, NTT Laboratories, and NAOJ. We also acknowledge the referee, Albert Zijlstra, for careful reading and providing several fruitful comments for paper improvement.  HI and SD have been financially supported by Grant-in-Aid for Scientific Research from Japan Society for Promotion Science (20540234). JN was supported by the Research Grants Council of the Hong Kong under grants HKU703308P, and by the financial support from the Seed Funding 
Programme for Basic Research in HKU (200802159006).

 \clearpage


 \begin{table*}[h]
 \caption{Parameters of the water fountain sources and the ASTE observations.}
 \label{tab:water-fountains}
 \begin{tabular}{llrrrcrlrcr}
 \\  \hline  \hline
 & & $l$\footnotemark[1] & $V_{\mbox{\scriptsize sys}}$\footnotemark[2] 
& $  \Delta V_{\mbox{\scriptsize los}}$\footnotemark[3] 
& $D$\footnotemark[4] & $t_{\mbox{\scriptsize jet}}$\footnotemark[5] & 
& Dur\footnotemark[7] & On-\footnotemark[8] & rms  \\
IRAS name & Other name & (\arcsec) &  \multicolumn{2}{c}{(km~s$^{-1}$)} 
& (kpc) & (year) & Ref.\footnotemark[6] & (hr) & point & (mK)  \\  \hline
15445$-$5449 & OH~326.5$-$0.4 & & $\sim -100$\footnotemark[9] & 111 & $  \sim$7 & ? & 2 & 0.90 & 1 &  10 \\
15544$-$5332 & OH~325.8$-$0.3 & & $\sim -74$\footnotemark[9] & 40 & 3.5 or 10.4\footnotemark[11] & ? 
& 2 & 0.90 & 5 &  62\\
16342$-$3814 & OH~344.1$+$5.8 & 2.4 & 42$\pm$2 & 240 & 2.0 & $ \sim$100 & 1  & 4.15 & 1 & 4 \\
16552$-$3050 & GLMP~498 & 0.11 & $\sim$12\footnotemark[10] & 170 & 19.6\footnotemark[11] 
& $\sim$60 & 8, 9 & 0.75 & 1 & 8 \\
18286$-$0959 & & 0.24 & 41$\pm$1 & 200 & 3.1 & $ \sim$15 & 3, 4 & 2.10 & 5, 9 & 35 \\
18450$-$0148 & W43A, OH31.0$+$0.0 & 0.8 & 34$\pm$1 & 190 & 2.6 & 50 & 6, 7 & 1.10 & 5 & 20 \\
18460$-$0151 & OH~31.0$-$0.2 & 0.11 & 125$\pm$1 & 290 & 2.0 & $ \sim$5 & 3, 4 & 1.00 & 9, 1 & 34 \\
18596$+$0315 & OH~37.1$-$0.8 & & 90$\pm$1 & 59 & 4.6 or 8.8\footnotemark[11]  & ? & 2 & 0.90 & 1 & 19  \\
19134$+$2131 & & 0.15 & $-$65$\pm$2 & 100 & 8.0 & 40 & 5 & 1.50 & 1 &  4 \\  \hline
 \end{tabular}

 \noindent
 \footnotemark[1]Total angular length of the jet system.  \\
 \footnotemark[2]Systemic velocity of the jet system.  \\
 \footnotemark[3]Full range of the line-of-sight velocities of water maser emission.  \\
 \footnotemark[4]Distance to the source.  \\
 \footnotemark[5]Dynamical age of the jet 
($ \approx l/ \Delta V_{\mbox{\scriptsize los}}$).  \\
 \footnotemark[6]References of the jet parameters. 
1:  \citet{cla09}; 2:  \citet{dea07}; 3:  \citet{deg07}; 4: \citet{ima07b}; 5:  \citet{ima07}; 6:  \citet{ima05}; 7:  \citet{ima02};  8:  \cite{sua07}; 9: \citet{sua08}  \\
 \footnotemark[7]Duration of the total observation time with ASTE.  \\
 \footnotemark[8]Number of points observed on and around the target, except an off-point.  \\
 \footnotemark[9]Both of the OH and \h2o maser spectra have either irregular or single-peaked profiles, causing difficulty in estimating  a precise systemic velocity. \\
 \footnotemark[10]No OH maser is detected and the \h2o maser spectrum has an irregular profile, causing difficulty in estimating  a precise systemic velocity. \\
 \footnotemark[11]Kinematic distance estimated on the basis of the Galactic kinematical parameters provided 
 by \citet {rei09} and \citet{sof09} is given. 
 \end{table*}

 \clearpage

 \end{document}